\Title page:

# High pressure superconducting phase diagram of $^6$Li: anomalous isotope effects in dense lithium

Authors: Anne Marie J. Schaeffer, Scott R. Temple, Jasmine K. Bishop and Shanti Deemyad

Affiliations:

Department of Physics and Astronomy, University of Utah

Contact Information:

[#]Correspondence to: deemyad@physics.utah.edu

**\Significance:**

The emergence of exotic quantum states, such as fluid ground state and two component superconductivity and superfluidity, in a compressed light metallic system has been entertained theoretically for metallic phases of hydrogen. The difficulty of compressing hydrogen to metallization densities, has prevented experimental proof of these effects. Studying lithium, which is isovalent to hydrogen and the lightest metal, is considered as a route to studying the lattice quantum effects in a dense light metallic system. Here, by comparing the superconductivity of lithium isotopes under pressure, we present evidence that properties of lithium at low temperature are significantly dominated by its lattice quantum dynamics. This study is the first experimental report on superconducting properties of $^6$Li; the lightest superconducting material.

\Abstract

We measured the superconducting transition temperature of $^6$Li between 16-26 GPa, and report the lightest system to exhibit superconductivity to date. The superconducting phase diagram of $^6$Li is compared to that of $^7$Li through simultaneous measurement in a diamond anvil cell (DAC) (1, 2). Below 21 GPa, Li exhibits a direct, but unusually large isotope effect, while between 21-26 GPa, lithium shows an inverse superconducting isotope effect. The unusual dependence of the superconducting phase diagram of lithium on its atomic mass provides evidence that lattice quantum dynamic effects dominate the low temperature properties of dense lithium.

\Body

Introduction

Light elements (Low Z) and their compounds have been the subject of many studies for their potential as high temperature superconductors. Due to the significant zero point effects, low Z compounds have large lattice dynamics. The properties of low Z materials, therefore, may deviate from the predictions made by static lattice models. Dependence of superconductivity on isotopic variations of these compounds can be used to probe and determine the magnitude of such effects.

Under ambient pressure, lithium is the lightest metallic and superconducting system and it exhibits one of the highest superconducting transition temperatures of any elemental superconductor under compression (3-7). Despite the large mass difference between the stable isotopes of lithium (~15%), isotope effects in superconductivity of lithium have not been studied before.

In systems that have long range interactions, such as metals, the ratio of lattice zero-point displacements to interatomic distances may increase under compression, provided they retain their long range interactions (8, 9) (This is opposite of systems with short range interactions, e.g. helium, in which the lattice becomes more classical under compression). In these systems, more deviations from the static lattice behavior are expected at higher densities. The significance of lattice quantum dynamics in lithium, even at ambient pressure, is evident from its temperature driven martensitic transition from bcc to hR9. At sufficiently low temperatures, where thermal energy is small, lattice quantum dynamics can play a more dominant role in the bulk properties. Sound velocity measurements on stable isotopes of lithium at 77 K and up to 1.6 GPa show that quantum solid effects in lithium, at least in the pressure range studied, do not decrease as a function of pressure(10). Raman spectroscopy studies between 40-123 GPa and at 177 K report a reduced isotope effect in high frequency vibrational modes of Li, which may be related to quantum solid behavior(11). Up to this point, no experiments have reported a comparison of any physical properties of lithium isotopes at low temperatures and high pressures concurrently. Since the superconducting transition of lithium occurs in a relatively low temperature range (12-14), studying its superconducting isotope effect provides excellent conditions to search for quantum lattice effects and their evolution as a function of pressure.

In the present study, we have measured the superconducting isotope effects in the stable isotopes of lithium under pressure. Lithium is a simple metallic system which is expected to exhibit conventional phonon mediated superconductivity and a well-defined superconducting isotope effect with nominal pressure dependence(15). Since phonon-mediated superconductivity depends on lattice and electronic properties of a material, any unusual isotopic mass dependence of the superconducting phase diagram can be indicative of the effects of large lattice quantum dynamics on electronic and/or structural properties.

**Experiment and results**

The expected conventional isotopic change in T$_c$ of lithium is $T_c^6 \approx 1.08 T_c^7$, which leads to a relatively small temperature shift at low temperatures, where lithium becomes a superconductor ($T_c^7(max) \leq 20K$, for which a BCS isotope shift of +1.6 K is expected for $^6$Li) (12-14). This small difference in the relative T$_c$'s and lack of *in situ* thermometry in a DAC make conducting experiments sensitive enough to resolve the differences challenging. Any inconsistency in the temperature measurement between experimental runs may mask the expected isotope effect. In addition to experimental uncertainties, differences in a sample's thermal history may change T$_c$(16). This is especially so for Li, in which the boundaries of the martensitic transition can be shifted if the sample is not annealed(17). To achieve the required resolution in evaluating the relative T$_c$'s of the two samples, both isotopes were measured simultaneously inside the same DAC. The details of these experiments are given below, and the general principles of the method used have been previously published(1) (fig. 1). Previous comparative measurements on the low temperature electrical resistivity on lithium isotopes under ambient pressure also noted the importance of simultaneous measurements(16); however, simultaneous electrical measurements on large samples under ambient pressure are less technically limiting than similar measurements under high pressure. Considerations regarding the thermal history are important for any comparative studies, such as structural or magnetic studies on lithium isotopes.

In the present work, we have used two isotopically rich samples of lithium; a $^6$Li-rich sample (the $^6$Li samples contained 99.99% lithium with the isotopic composition 95.6% $^6$Li and 4.4% $^7$Li together with the impurities Na, Mg, Al, and other elements; Sigma Aldrich) and a $^7$Li-rich sample (the isotopic composition of the 99.9% pure natural lithium ($^7$Li) was 92.41% atomic $^7$Li and **7.**59% atomic $^6$Li; the impurity composition was the same as in the case of the ⁶Li; Sigma Aldrich), which we refer to as $^6$Li and $^7$Li samples respectively for the remainder of this manuscript.

All measurements were carried out in a DAC using electrical resistance as a means of determining the superconducting transition temperatures. All pressure increases were carried out at room temperature. The return to room temperature after every measurement allowed the samples to be transformed to their equilibrium phase. Lithium reacts readily with many materials including diamond (12, 14, 18), which may cause a DAC to fail. In order to prevent any such reactions, compressed, dehydrated alumina powder, which was heat treated at 110°C to remove moisture, was used both as an electrical insulator and a pressure medium(19). This allowed us to keep the samples at room temperature inside the DAC without risking failure of the diamonds. An insulated gasket was made from a 250 µm thick stainless-steel foil, pre-indented to a thickness of 40-55 µm. Two pressure chambers with initial diameters of 110 µm were symmetrically drilled 10-20 µm apart from each other on the gasket and several ruby spheres were dispersed evenly in each pressure chamber. The gasket then was insulated with a mixture of epoxy and alumina.

The use of alumina as pressure medium exposes the sample to non-hydrostatic conditions that may in principle affect the superconductivity. However, in the case of lithium, studies up to 50 GPa in which no pressure medium was used, show very sharp Bragg peaks. This provides evidence that lithium itself remains a very soft solid which does not support large shear stresses (20, 21). (The superconducting phase diagram of $^7$Li measured using helium as hydrostatic pressure medium by Deemyad and Schilling and non-hydrostatic measurements by Struzhkin et. al. without any pressure medium, are very similar below 30 GPa (figure 3 c)). In the current work, the presence of quasi-hydrostatic conditions is supported by the sharpness of ruby peaks.

We have used high precision spectroscopy to measure the pressure distribution of each chamber from several ruby spheres. The maximum pressure gradient remained below ±0.8 GPa up to the highest pressure measured (See Fig 1. and the supplementary materials for further discussion). In a twin chamber gasket, not only the absolute pressure difference between the chambers and the pressure gradient across each sample is measured but also each chamber on its own acts as an independent indicator of the pressure dependence of the properties of its sample.

An isolated quasi-four probe arrangement was built on each sample chamber using platinum electrodes. To eliminate the interference between the two circuits, each circuit was connected to a separate lock-in amplifier and run on a different frequency (~ 5 and 13 Hz). To prevent any possible chemical reactions, the samples were loaded and pressurized inside a high purity argon glovebox. An AC current of $I_{rms}$~100µA was applied across each sample. Since the arrangement used here was a quasi-four probe, a small portion of the signal always came from the piece of Pt electrode in the path (fig. 1). The onset of superconductivity was defined by the temperature at which the resistance of sample drops to zero (fig. 2a). As an additional test of superconductivity, we used a small magnetic field (~100 Oe) to suppress $T_c$ (fig. 2b and 2c). Fig. 3a shows the superconducting phase diagram of $^6$Li and $^7$Li for pressures between 16 and 26 GPa. The experiments were completed with 6 separate loadings of the lithium isotopes, all of which overlapped in pressure range and showed complete internal consistency.

Fig. 3a shows the superconducting phase diagram of $^6$Li and $^7$Li for pressures between 16 and 26 GPa. The correlation between the superconducting transition temperatures of the two isotopes is anomalous and in the range of this study three distinct regions can be identified. The sign of $\left(\frac{dT_c}{dP}\right)$ in the pressure range $16 GPa < P < 26 GPa$ changes two times for $^6$Li. In this region, the slope is always positive, though not constant, for $^7$Li. A change in slope may be indicative of the presence of a structural phase boundary (such as hR9 to fcc for $^7$Li). Between 16 and 21.5 GPa, the superconducting $T_c$ of $^6$Li is higher than that of $^7$Li. Figure 3c shows the calculated values of the superconducting isotopic coefficient, $\alpha$ ($T_c \propto M^{-\alpha}$.), as a function of pressure. The lowest pressure points at 16 and 18 GPa for $^6$Li display an initial positive slope $\left(\frac{dT_c}{dP}\right)$, however, the paucity of data in this region does not allow to properly assign a slope of $\alpha$ vs. P. For $18 GPa < P < 21.5 GPa$, the value of the isotopic coefficient decreases monotonically with pressure, and the value of $\alpha$ is always higher than expected for BCS-type superconductors, $1 \lesssim \alpha \lesssim 4$. For $P > 21.5\ GPa$, the value of the isotopic coefficient, $\alpha$, changes sign and remains constant within error ($\alpha \approx -1.5 \pm 0.5$).

Fig. 3b shows the superconducting phase diagram of $^7$Li which has been plotted here together with all the previous measurements. The current result only overlaps with the studies of Deemyad and Schilling for 22 GPa < P < 26 GPa. As shown in figure 3c, the measurements of the superconductivity of natural lithium in the present work display the same trend as Deemyad and Schilling (with the shallow slope at the beginning followed by rapid increase in the slope). The present measurements consistently show slightly lower transition temperatures than Deemyad and Schilling, which can be caused by differences in thermometry and/or thermal histories of the samples and other differences between the methods that have been used. For example, Deemayd and Schilling never annealed their samples above 100 K, which, according to the phase diagram suggested by Guillame et al. may not allow the sample to relax to fcc or bcc phase between pressure applications below 20 GPa. In the present experiments, we applied the pressures at room temperature. For the comparison of the transition temperatures of the two isotopes, however, the samples compared need to have experienced the exact same thermal history and their relative transition temperature must be known precisely. All previous measurements of the high pressure phase diagram of lithium done in past (excluding the measurements of the superconducting phase diagram of natural lithium by Deemyad and Schilling) used no pressure medium (11, 12, 14, 18, 20-22). The experiments done by Deemyad and Schilling used helium as pressure medium and found very reproducible results. The results, however, may have been influenced by diffusion of helium into the lithium lattice. Moreover, Deemyad and Schilling as well as Struzhkin et al. used magnetic susceptibility to detect the superconducting transition. Magnetic susceptibility measurements currently cannot be employed for simultaneous measurements; critical for an experiment designed to characterize the slight differences between isotopes.

**Discussion**

For a BCS-type superconductor composed of one atomic species, with a static lattice, the superconducting transition temperature ($T_c$) and the ionic mass follow the simplified relationship, $T_c \propto M^{-\alpha}$, in which $\alpha$=0.5. This relation is mainly attributable to differences in the Debye temperature for different ionic masses of different isotopes. In the current experiment, we observe an anomalous isotope effect in the superconducting phase diagram of lithium that cannot be explained by conventional models used for solids with static lattice.

In a non-static lattice and in the presence of large quantum zero point motion of ions, electrons do not see the lattice as a perfect crystal (this is the case even if these displacements do not affect the structures of the isotopes) thus, in total, drastic deviations from conventional isotope effects in a superconducting quantum solid can be expected(23, 24).

The large departure from BCS-type behavior may also be explained if the two isotopes have different structures in this pressure range. For a solid with static lattice and in the absence of large zero point energy, the equilibrium distance between the lattice particles is determined by the minimum of the potential energy of the lattice and, to first approximation, is independent of the particles' mass (25, 26) and isotope effects in the structures are not expected. On the other hand, in a solid with large lattice quantum dynamics, the large zero point energy of the lattice will significantly contribute to the vibrational and rotational energies, which can have an impact on the equilibrium structures (25, 26). Since BCS assumes identical structures for isotopes, the BCS superconducting isotope relation is not applicable in systems with a structural isotope effect. In the case of lithium, it has been theoretically shown that without inclusion of zero point energy some of the known structures of natural lithium (oC88) will not be stable. Therefore, it is not unexpected if the differences in the zero point energies of lithium isotopes lead to different structural phase diagram for them (27).

The structures of Li as a function of pressure (for pressures above 0.65 GPa(28)) are not known for T < 77 K, which contains the boundaries of the martensitic phase(18). This phase is thought to play an important role in superconductivity. Moreover, the isotopic dependence of the structures of lithium is not known in this pressure regime. Between 16-22 GPa, $^6$Li exhibits a change from a positive to a negative slope in $T_c$. In contrast, $^7$Li shows a positive slope in the entire range where superconductivity was observed. It is possible that the large difference in $T_c$'s, the initial change in slope for $^6$Li, and the opposite slopes observed between 18-22 GPa are caused by a low temperature structure in $^6$Li not present in $^7$Li. Low temperature structural studies are required to characterize this region. Above ~21 GPa, the superconducting $T_c$ of $^6$Li falls below that of $^7$Li, this effect remains until the highest pressure studied here, ~26 GPa. The inverse isotope effect in conventional superconducting systems is noted in some transition metal hydrides (MH), such as PdH, and in the element $\alpha$-U (29-32). The inverse isotope effect in MH has been attributed to the existence of large quantum effects in hydrogen, which cause anharmonicity in phonon spectra(33, 34); an explanation which cannot account for the behavior of $\alpha$-U. If the unusual isotope effect in Li is solely a consequence of its zero point displacements and a consequent anharmonicity in its lattice, Li is the only elemental solid known to exhibit such behavior. However, in the present case the electronic and structural effects may be entangled. Comparative structural studies for lithium isotopes at low temperature are very challenging but they are currently possible in 3rd generation synchrotron sources. The results of this study would justify the investment on such works. The detailed comparative studies on additional low temperature properties of lithium isotopes may also shed light on the predicted properties of metallic hydrogen.

It is also notable that isotopes of Li possess different quantum statistics, and natural lithium has been shown to exhibit nuclear order at one atmosphere(35). The magnetic order in lithium isotopes may contribute to their superconducting isotope effects at ambient pressure where natural lithium superconducts below 0.4 mK. Studying the isotope effects in ambient pressure superconductivity of lithium would be also enlightening.

## Acknowledgments

Authors are grateful for experimental assistance from R. McLaughlin, D. Sun, H. Malissa, Z. Jiang, F. Doval and A. Friedman . This work is supported by NSF-DMR grant # 1351986. JKB acknowledges the financial support from the University of Utah UROP fund.

\Figures:

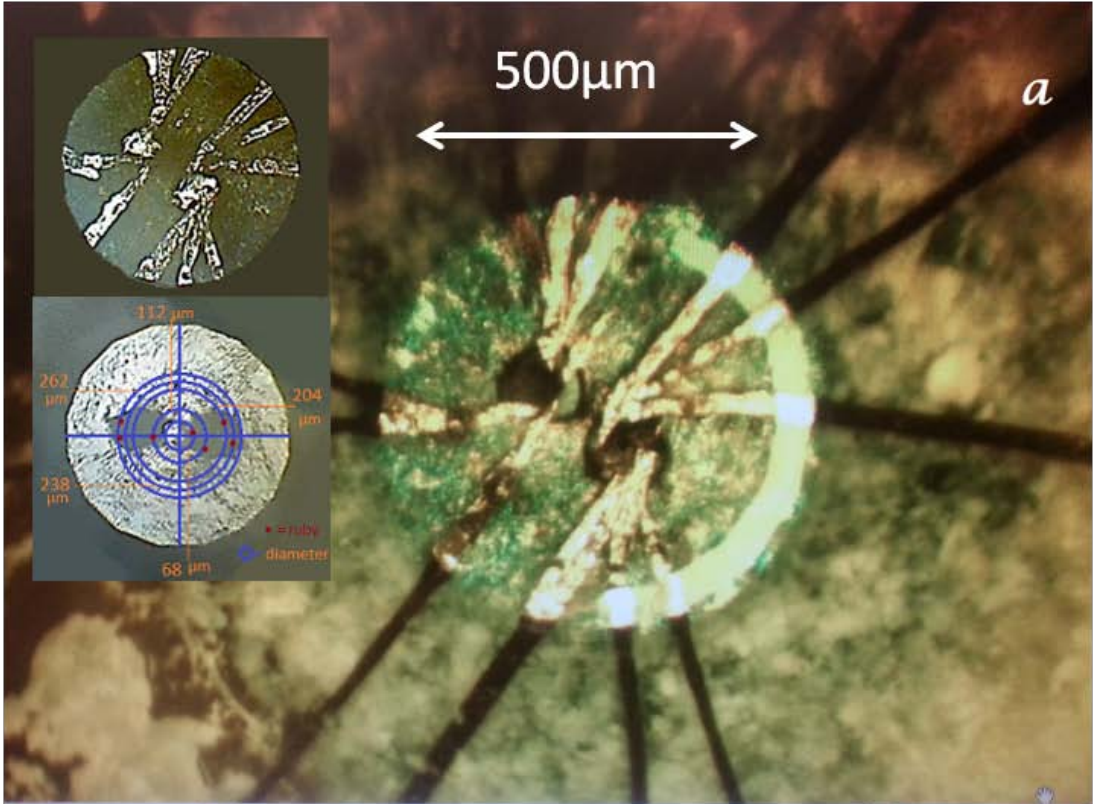

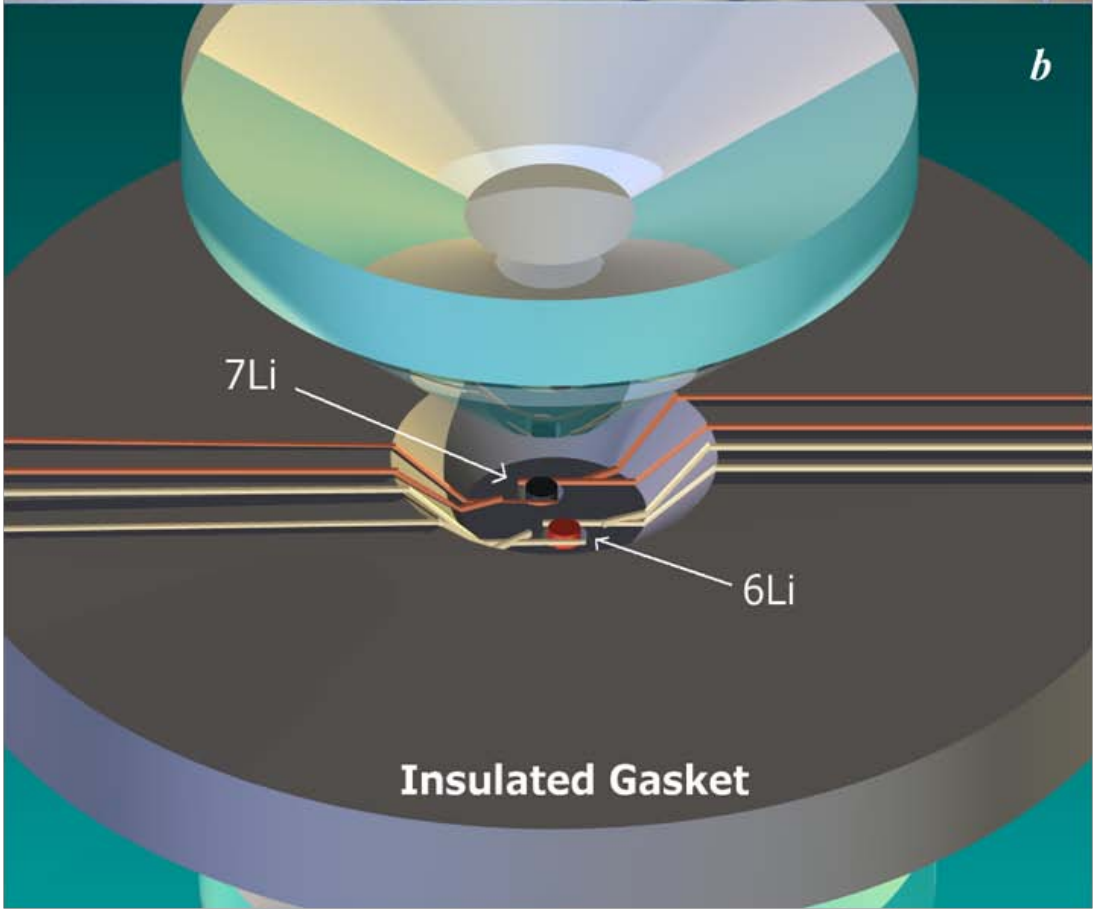

**Figure 1)** a) Twin chamber gasket built on a 500 μm culet diamond which is used in the present experiments in a DAC for simultaneous measurements of superconductivity. Each pressure chamber has a pair of extra Pt leads and contains several pieces of ruby for accurate determination of pressure gradient within each pressure chamber. The insets show the gasket and samples under reflected light, at ~21 GPa, demonstrating the metallic appearance of both samples and map of ruby pieces inside each pressure chamber in the same run . b) Schematic drawing of the twin chamber design used in the experiments. Small portions of the platinum leads in the path contribute to the total resistance measured for each sample. The electrodes for measuring the resistance of $^6$Li and $^7$Li are shown in different colors.

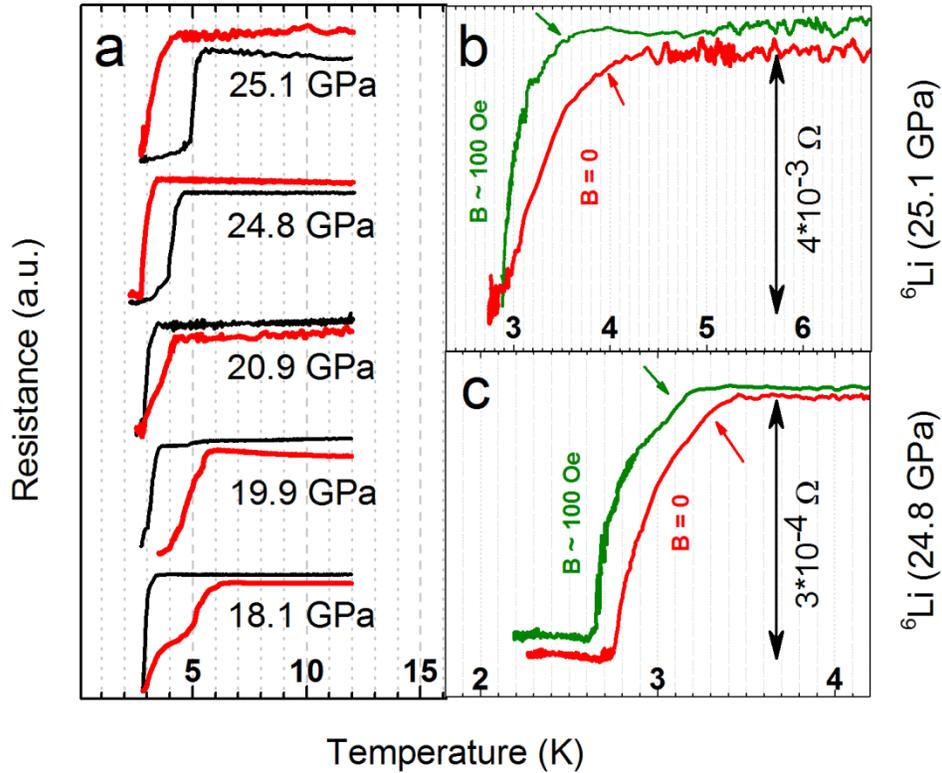

**Figure 2)** a) Superconducting transitions as determined by electrical resistivity. All black lines show transitions for $^7$Li, red lines are the transitions for $^6$Li. Each pair shows the simultaneous measurement. The double step in 18.1GPa transition of $^6$Li may be related to presence of mixed phases with different T$_c$'s at a structural phase boundary. The samples' resistances in a normal state above their superconducting transitions is ~0.5-10mΩ varying by the sample size and geometry. These values would give an estimate of $\rho \approx 0.5 - 1\mu\Omega.cm$, at room temperature, for a typical sample size of $50 \times 50 \times 10 \mu m^3$. An RRR value of ~75 is estimated from ambient pressure measurements on the samples used here (22). The transitions above are scaled for ease of comparison. b,c) The shift of T$_c$ with an applied external magnetic field of B~100 Oe for $^6$Li at 23.3 and 26.6 GPa. B~100 Oe for green lines and B=0 for red lines.

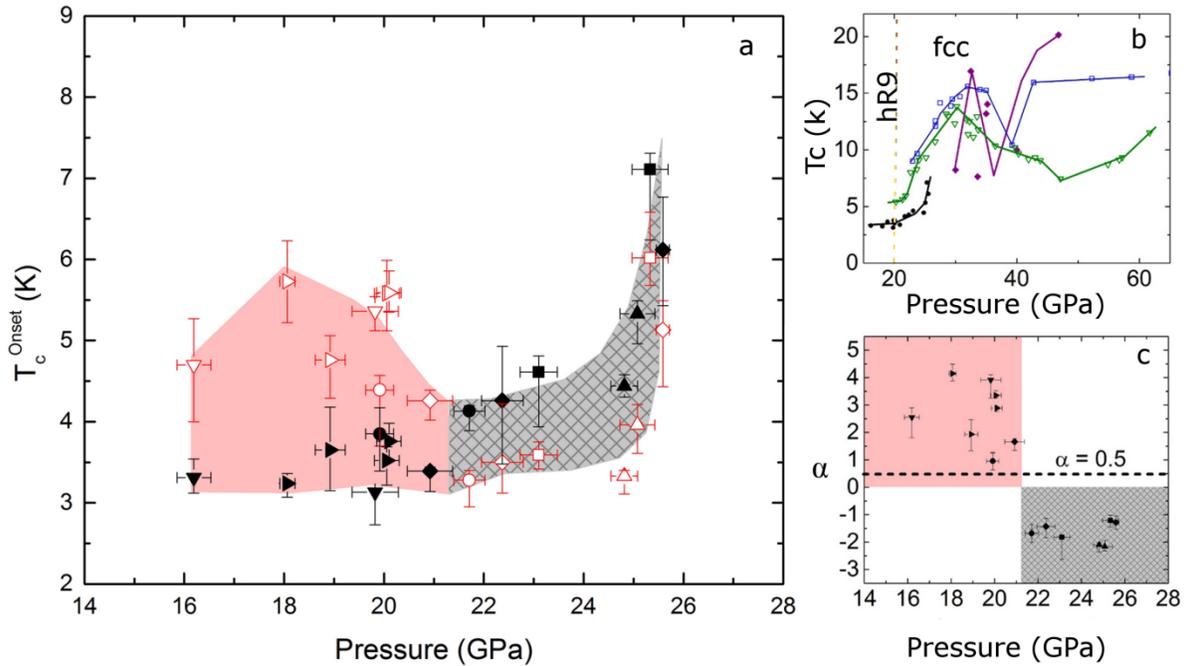

**Figure 3)** a) Superconducting phase diagram of lithium isotopes. Open shapes represent $^6$Li, solid shapes represent $^7$Li. The different shapes designate separate loadings. The pink shaded region from 18 GPa to 21.5 GPa shows the direct isotope effect with a large difference in the Tc's for $^6$Li and $^7$Li. The grey shaded region shows the inverse isotope effect from 21.5 GPa to 26 GPa. The pressure error bars represent the maximum pressure difference between all the rubies shown in figure 1 in the two chambers (The value is generally is equal to the difference between the pressure from the ruby in the smallest radii of one chamber and the ruby in the largest radii of the opposite chamber). b) Comparison of the various superconducting phase diagram of natural lithium measured by various techniques (Open triangles: Deemayd and Schilling, Open squares: Struzhkin et al., diamonds: Shimizu et el. and circles: This study). The solid lines are guide to eye. Dashed line is the speculated boundary between hR9 and fcc at low temperature. c)The isotope coefficient, $\alpha$, as a function of pressure. The dashed line at $\alpha = 0.5$ shows the expected value for a conventional isotope effect.

\Supplementary materials

Use of solid pressure medium is currently inherent to electrical measurements in diamond anvil cells and the reactivity of lithium with most pressure media, including diamond itself, limits the choice of materials used for a pressure medium. To further analyze the pressure gradient in the twin chamber design with alumina as pressure medium we ran series of experiments on a twin chamber gasket filled with alumina powder and packed each chamber with large number of ruby spheres and compared the pressure gradient between the chambers. We used the 500 µm culet diamonds, as in the manuscript, as well as the same chamber dimensions that we used in the experiments of the present work. Using a high resolution ruby microscope, we measured the pressure gradient across both chambers. The graph below shows the correlation between the pressures of the rubies in approximately symmetric positions in the two chambers. The comparison between the pressures of the two chambers shows a correlation of 1 with relatively little spread.

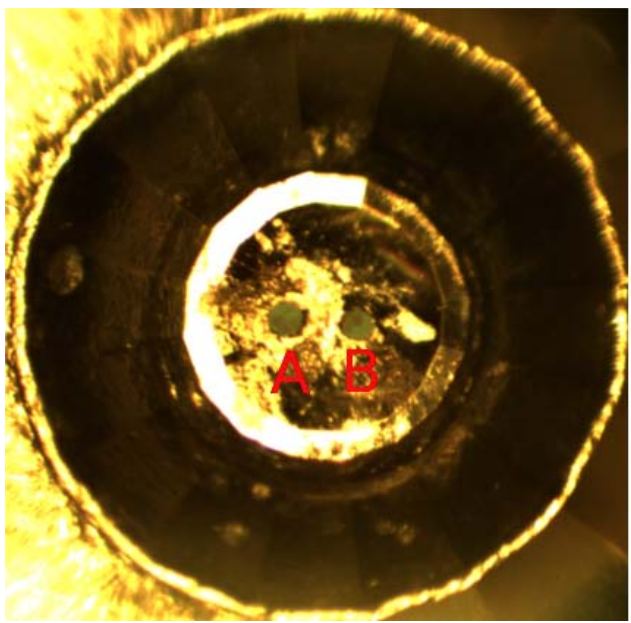
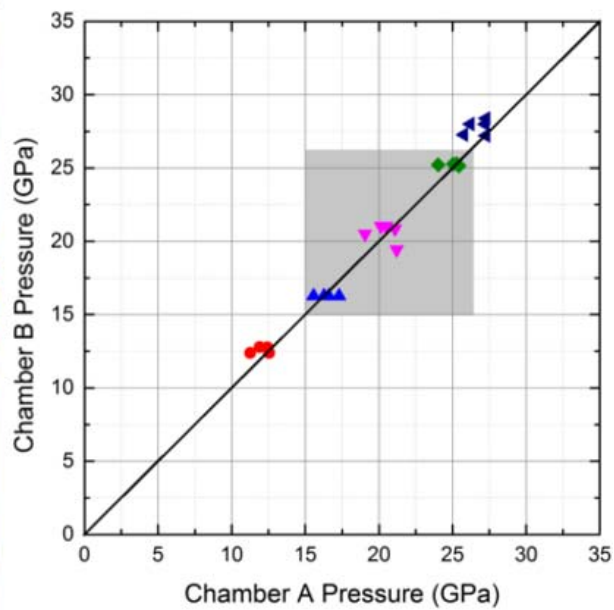

The above photos show the twin chamber gasket lit by reflected. On the right side the pressure measured from rubies in chamber A were compared to rubies in chamber B for the rubies that fell on approximately the same radii.

The graph below shows the pressure distributions in each chamber. The horizontal axis is the distance of each ruby from the center of the gasket. Each pressure increase has a different symbol. Solid and open symbols represent pressure from the rubies in chamber A and B respectively. Shaded region correspond to the pressures in the manuscript.

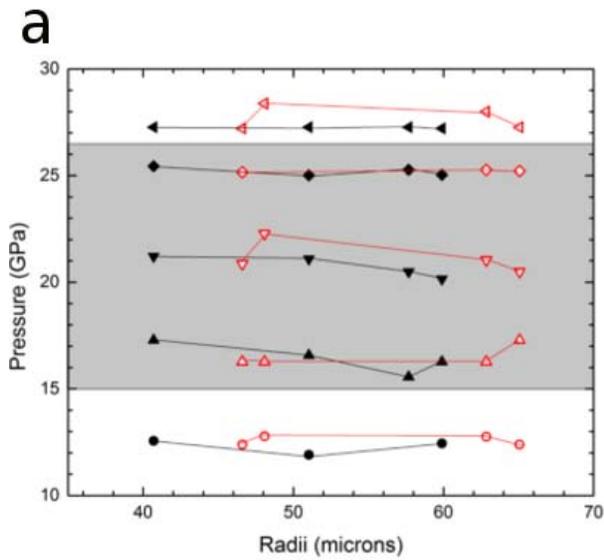
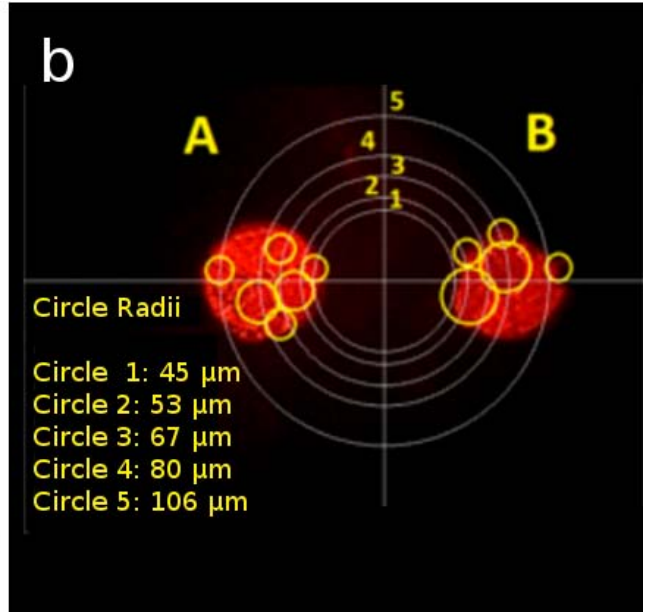